\documentclass[prb,twocolumn,amsmath,amssymb,floatfix,superscriptaddress]{revtex4-1}
\usepackage{graphicx} 
\usepackage{booktabs}
\usepackage{color}
\usepackage[version=4]{mhchem}
\usepackage{xspace, bm, braket}
\usepackage{siunitx}
\usepackage{verbatim}
\usepackage{subfigure}
\usepackage[colorlinks, citecolor=blue, linkcolor=blue, urlcolor=blue, breaklinks]{hyperref}
\usepackage{placeins}
\bibpunct{[}{]}{,}{n}{}{}

\newcommand{\SRO}{\ce{Sr2RuO4}}
\newcommand{\SMO}{\ce{Sr2MoO4}}
\newcommand{\BOO}{\ce{BaOsO3}}

\newcommand{\imag}{\text{i}}
\newcommand{\angstrom}{\textup{\AA}}

\usepackage[svgnames]{xcolor}
\newif\ifshowcomments\showcommentstrue
\def\beq{\begin{equation}}
\def\eeq{\end{equation}}

\begin{document}
	
	\title{BaOsO$_3$: A Hund's metal in the presence of strong spin-orbit coupling}
	
	\author{Max Bramberger}
	\email{M.Bramberger@physik.uni-muenchen.de}
	\affiliation{Arnold Sommerfeld Center of Theoretical Physics, Department of Physics, University of Munich, Theresienstrasse 37, 80333 Munich, Germany}
	\affiliation{Munich Center for Quantum Science and Technology (MCQST), Schellingstrasse 4, 80799 Munich, Germany}
	\author{Jernej Mravlje}
	\affiliation{Jo\v zef Stefan Institute, Jamova 39, Ljubljana, Slovenia}
	\author{Martin Grundner}
	\affiliation{Arnold Sommerfeld Center of Theoretical Physics, Department of Physics, University of Munich, Theresienstrasse 37, 80333 Munich, Germany}
	\affiliation{Munich Center for Quantum Science and Technology (MCQST), Schellingstrasse 4, 80799 Munich, Germany}
	\author{Ulrich Schollw\"ock}
	\affiliation{Arnold Sommerfeld Center of Theoretical Physics, Department of Physics, University of Munich, Theresienstrasse 37, 80333 Munich, Germany}
	\affiliation{Munich Center for Quantum Science and Technology (MCQST), Schellingstrasse 4, 80799 Munich, Germany}
	\author{Manuel Zingl}
	\affiliation{Center for Computational Quantum Physics, Flatiron Institute, 162 5th Avenue, New York, NY 10010, USA}
	
	\date{\today}
	
	\begin{abstract}
		We investigate the 5d transition metal oxide BaOsO$_3$ within a
		combination of density functional theory (DFT) and dynamical mean-field
		theory (DMFT), using a matrix-product-state impurity solver. BaOsO$_3$ has 4 electrons in the  t$_{2g}$ shell akin to ruthenates but stronger spin-orbit coupling (SOC)
		and is thus expected to reveal an interplay of Hund's-metal behavior
		with SOC.  We explore the paramagnetic phase diagram as a function
		of SOC and Hubbard interaction strengths,
		identifying metallic, band (van-Vleck) insulating and Mott
		insulating regions. At the physical values of the two couplings
		we find that BaOsO$_3$ is still situated inside the metallic region and has a
		moderate quasiparticle renormalization $m^*/m \approx 2$; consistent with specific heat measurements. SOC leads to a splitting of a van-Hove singularity (vHs) close to the Fermi energy and a subsequent reduction of electronic correlations (found in the vanishing SOC case), but the SOC strength is insufficient to push the material into an insulating van-Vleck regime. In spite of the strong effect of SOC, BaOsO$_3$ can be best pictured as a moderately correlated Hund's metal.
		
	\end{abstract}
	
	\maketitle
	
	\section{Introduction}
	The fascinating properties exhibited by many quantum materials are often the result of
	a complex interplay of several factors, such as the on-site Coulomb repulsion $U$, the inter-orbital Hund's coupling $J_H$, and band structure details like crystal-fields and van-Hove singularities (vHs)~\cite{imada98,Georges2013}. Additionally, at large atomic
	numbers spin-orbit coupling (SOC) starts to play an increasingly important
	role~\cite{Kim2008,Martins2017,Triebl2018}, which can lead to the occurrence
	of Mott insulating states~\cite{Kim2008,Martins2011,Arita2012, meetei2015,buenemann2016} and
	also to nontrivial topology~\cite{Pesin2010,zhang2017}.
	
	A family of quantum materials that has been given a lot of attention recently are
	Hund's metals, which are characterized by the inter-orbital Hund's coupling $J_H$ governing electronic correlations~\cite{Yin2011, medici11,Georges2013,Lanata2013,fanfarillo15,Stadler2015,Stadler2019,Horvat2019}.
	In the literature on Hund's metals the main focus has been the study of 3d and 4d transition metal oxides, like the iron-pnictides or the ruthenates. An important Hund's metal where the combination of experimental and theoretical
	investigations has led to an unprecedented understanding of
	nontrivial correlated behavior is 
	\SRO{}~\cite{Mackenzie1996b,Maeno1997,Bergemann2003,Stricker2014,
		Behrmann2012, Tamai2018, Deng2016, Veenstra2014,
		Zingl2019,Sarvestani2018,Lee2020,Zhang2016,Kim2018,Strand2019,Aiura2004,Mravlje2011,Deng2019,Lee2020,Kugler2020, Karp2020}.
	Besides Hund's physics, a key role in terms of the surprisingly strong correlations in \SRO{} was attributed to the presence of a vHs in close vicinity of the Fermi energy~\cite{Mravlje2011,Lee2020,Kugler2020,Karp2020}.
	
	The role of the SOC was explored too~\cite{Haverkort2008,Veenstra2014,Tamai2018,Liu2008,Zhang2016,Kim2018,Tamai2018,Linden2020}.
	It was found that SOC does not affect the quasiparticle mass enhancement appreciably, but has important consequences on the detailed shape of the Fermi
	surface~\cite{Haverkort2008,Behrmann2012,Veenstra2014}, and is subject to an effective correlation-driven enhancement by as much as a factor of two~\cite{Liu2008, Zhang2016,Kim2018,Tamai2018,Linden2020}.
	
	However, even in the 4d ruthenates, the bare SOC is still at moderate \SI{0.1}{eV}, which is 3-4 times smaller than the Hund's coupling. It is an open question to what extent the intriguing physics found in 3d/4d Hund's metals survives in compounds that have a substantially stronger SOC.
	
	Iridates, for example Sr$_2$IrO$_4$ and NaIrO$_3$, are 5d materials with a SOC of about \SI{0.3}{eV} where electronic correlations are known to play an important role ~\cite{Lenz2019,Arita2012,Yamasaki2014,Li2013,Zhang2013,Martins2011,Bhandari2019,Ye2013,Porras2019,Kim2012,Pincini2017,Gretarsson2016, du2013,Bremholm2011}.
	Sr$_2$IrO$_4$ is an insulator, since the SOC splits the $j$ states into fully occupied $j=3/2$ and half-filled $j=1/2$ states. In the latter the Hubbard repulsion is strong enough to drive the material into a Mott-insulating state. NaIrO$_3$ is also found to be an insulator, although it has a nominal Hund's-metal filling of 4 electrons in 3 orbitals. In this material the combination of SOC and the polarizing effect of the Hubbard repulsion is responsible for the formation of a band-insulating ground state. These examples raise the question if there are 5d compounds in which Hund's coupling is of similar importance as in the case of 4d materials. 
	
	A promising class of 5d oxides where the interplay of SOC, Hund's coupling and van-Hove physics can be studied is osmates. Although electronic correlations in osmates with a half filled t$_{2g}$ shell (NaOsO$_3$, LiOsO$_3$) have been given considerable attention~\cite{KimBong2016,Springer2020,Liu2020}, not so much is known on electronic correlations in osmates at a Hund's-metal filling.
	
	With 4 electrons in the three t$_{2g}$ orbitals, BaOsO$_3$ is a potential candidate for a 5d Hund's metal. This material crystallizes in a cubic perovskite structure and shows a metallic behavior in optical and transport measurements on polycrystalline samples~\cite{Shi2013,Zheng2014}. A specific heat enhancement of 2.2 over the band-structure value~\cite{Shi2013} indicates sizable electronic correlations for a 5d metal with quite spatially-extended correlated orbitals ($U$/bandwidth $\sim 0.8$). These observations suggest the classification of BaOsO$_3$ as a Hund's metal. 
	On the other hand, the large SOC of the order of the Hund's coupling ($\sim \SI{0.3}{eV}$) will introduce a tendency towards a band-insulating state like it was observed in the case of NaIrO$_3$~\cite{du2013,Bremholm2011}. This opens an interesting question: How does the Hund's rules desire to form a local high-spin compete with the SOC favoring a fully polarized state? Moreover, DFT calculations for \BOO{} (without SOC) showed that a vHs is present right at the Fermi level~\cite{Jung2014,Zahid2015}, which we expect to have a substantial influence on electronic correlations. Therefore, we believe that BaOsO$_3$ is an ideal candidate to gain a deeper understanding of the interplay of Hund's physics, SOC and van-Hove physics, as all these factors - and their complex interplay - are definitely relevant for the physics of this material.

	In this work, we shed light on the origin of electronic correlations in \BOO{}, using a combination of density functional theory (DFT) and dynamical mean-field theory (DMFT)~\cite{georges96,Kotliar2006} with a matrix product states (MPS)-based impurity solver~\cite{Linden2020,wolf15iii}. 
	At physical values of the interaction parameters and SOC we obtain a mass enhancement in perfect agreement with experiments. Our calculations show that electronic correlations in this compound are enhanced by two factors: the Hund's coupling and the proximity of a vHs to the Fermi energy. To understand the impact of these factors on electronic correlations in more detail, we explore a range of SOC and interaction strengths. Importantly, we find that correlations are diminished with respect to the case without SOC, because SOC removes the vHs from the Fermi level and weakens the effect of Hund's physics.
	
	We note that electronic correlations as a function of SOC and interaction strength were explored for the case of NaIrO$_3$ using the Gutzwiller approach~\cite{du2013}, for the Bethe lattice using both Gutzwiller and DMFT~\cite{du2013b}, and in the case of one dimensional model calculations using DMRG~\cite{kaushal2017}.

	The paper is structured as follows. In Sec.~\ref{sec:Method}, we discuss the used methods and the details of the calculations. In Sec.~\ref{sec:DFT} we present basic DFT results on \BOO{} with and without SOC and in Sec.~\ref{sec:spec} we investigate electronic correlations using DMFT. In Sec.~\ref{sec:vHs}, we explicitly show that the vHs plays a major role in this material. In order to better understand the interplay between Hund's coupling and SOC we study a variety of interaction parameters and SOC strengths in Sec.~\ref{sec:SOC}. Finally, in Sec.~\ref{sec:Phases} we map out a paramagnetic phase diagram in the plane of Hubbard repulsion and SOC, which puts our work in context with experimental observations.
	
	\section{Method}
	\label{sec:Method}
	We perform DFT calculations with Wien2k~\cite{Blaha2018} using the
	PBE-GGA~\cite{PBE} exchange-correlation functional and the experimentally determined $Pm\bar{3}m$ crystal structure with 
	$a = b = c = \SI{4.02573}{\angstrom}$~\cite{Shi2013}. The DFT calculations are converged on a $34 \times 34 \times 34 $ $k$-point grid with $RK_\mathrm{max}=7$. We use wien2wannier~\cite{wien2wannier} and Wannier90~\cite{wannier90} to construct maximally-localized Wannier functions~\cite{MLWF1, MLWF2} of t$_{2g}$ symmetry for the bands around the Fermi energy, employing a $10 \times 10 \times 10$ $k$-point grid together with a frozen energy window spanning the whole energy range of the t$_{2g}$ bands. We add SOC as a local t$_{2g}$-only term to the Wannier Hamiltonian, see e.g. Refs.~\cite{Zhang2016,Linden2020}. We neglect any possible $k$-dependence of the SOC.
	
	To consider electron-electron interaction within the t$_{2g}$-only Wannier model we include local interactions in Hubbard-Kanamori form~\cite{Kanamori1963}, as given in Appendix~\ref{app:method}.
	We believe that $U\sim \SI{2.55}{eV}$ and $J_H\sim \SI{0.27}{eV}$ are reasonable parameters to describe the physics of BaOsO$_3$. These values were reported in Ref.~\cite{Springer2020} for NaOsO$_3$ and LiOsO$_3$ and similar ones were found in Ref.~\cite{Ribic2014} for BaOsO$_3$. In addition to this `physical' parameter set we explore a wide range of interaction and SOC strengths.
	
	We solve the resulting multi-orbital Hubbard model within single-site
	DMFT~\cite{georges96} using the TRIQS library~\cite{TRIQS} and the TRIQS/DFTTools application~\cite{TRIQS/DFTTOOLS}. We use dense $51 \times 51 \times 51$ and  $101 \times 101 \times 101$ $k$-point grids for calculations with and without SOC, respectively. The calculations are ``one-shot" DFT+DMFT, meaning that the DFT density is kept fixed and not updated. The double counting correction is absorbed into the chemical potential, as we purely work in the low-energy subspace defined by the t$_{2g}$-like Wannier orbitals.
	
	All calculations are carried out with a MPS-based solver~\cite{wolf15iii,Linden2020} using the \textsc{SyTen} toolkit~\cite{hubig17, systen}. The MPS-based solver is similar to impurity solvers based on exact diagonalization~\cite{caffarel94} in the sense that it uses a discrete version of the Hamiltonian together with a finite bath. We emphasize that the MPS-based solver allows for the inclusion of SOC as well as calculations at zero temperature. Further methodological details on the MPS-based solver are provided in Appendix~\ref{app:mps} and Refs.~\cite{Linden2020,Karp2020,wolf15iii}. 

	We characterize the strength of electronic correlations by the inverse
	quasiparticle renormalization $Z^{-1}=1-\partial
	\mathrm{Im}\Sigma(i\omega_n\rightarrow 0)/\partial \omega_n$, equal to the quasiparticle mass enhancement $m^\star/m=Z^{-1}$ in the single-site DMFT approximation. We determine $Z$ by fitting a polynomial of 4th order to the lowest 6 points of the Matsubara self-energies and extrapolate $\mathrm{Im}\Sigma(\imag\omega_n\rightarrow 0)$, 
	a procedure also used in Refs.~\cite{Mravlje2011,Zingl2019,Karp2020}. The limit $\text{Re}\Sigma(i\omega_n\rightarrow 0)$ is also evaluated by fitting a 4th order polynomial to the first 6 Matsubara points, consistent with the computation of the mass enhancement.
	
	For real-frequency spectra we use TRIQS/maxent~\cite{TRIQS/maxent} to perform analytic continuations for the self-energy with the inversion method~\cite{Kraberger2017}. The renormalizations obtained by the polynomial fit of $\text{Im}\Sigma(\imag\omega_n)$ are in good agreement with the ones obtained with analytic continuation.

		\begin{figure*}[t]
		\centering
		\includegraphics[width=1.0\linewidth]{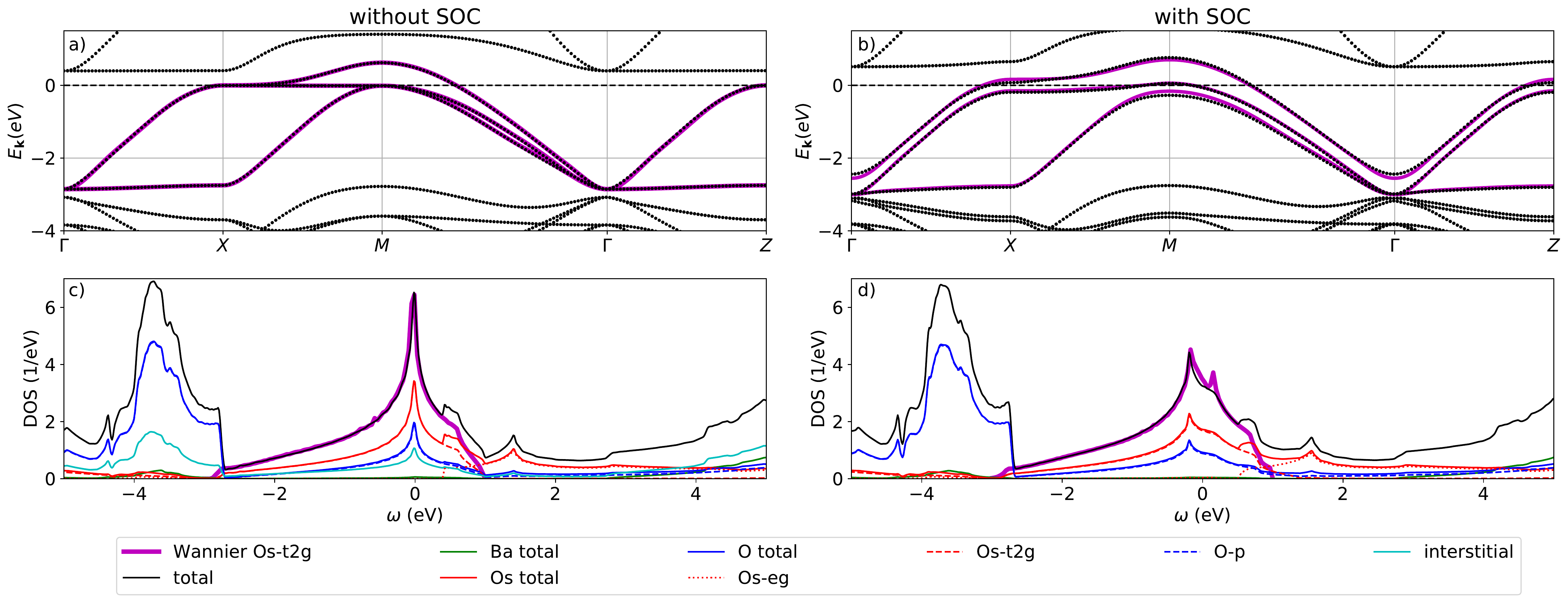}
		\caption{ Upper panels: DFT band structure (black dots) and Wannier fit (colored) of \BOO{} without (a)	and with (b) SOC along a high-symmetry path through the Brillouin zone. Lower panels: Total DFT DOS (black) and partial contributions (colored) without (c) and with (d) SOC. The DOS of the constructed $t_{2g}$-like Wannier orbitals is shown with thicker lines. For the case with SOC we have added a local SOC term to the Wannier Hamiltonian with a coupling strength of $\lambda=\SI{0.3}{eV}$. }
		\label{fig:dft_bands_dos}
	\end{figure*}
	
	\section{Results}
	We first discuss the physics of \BOO{} on the DFT level, which serves as a starting point for our DMFT calculations. We then study the effects of electronic correlations, including the competing phases in the vicinity of the physical parameter set for \BOO{}, as well as the influence of the SOC strength and the vHs on electronic correlations.
	
	\subsection{DFT and Wannier construction}
	\label{sec:DFT}

 	In \BOO{} the Os-d orbitals are responsible for the DOS around the Fermi energy, see Fig.~\ref{fig:dft_bands_dos}. These orbitals are well separated from the O-p states found at energies below about \SI{-2.5}{eV}. However, the octahedral oxygen environment surrounding the Os atoms leads to an e$_g$-t$_{2g}$ splitting of the Os-5d shell into empty e$_g$ orbitals and three degenerate t$_{2g}$ orbitals occupied by 4 electrons. The latter are strongly hybridized with oxygen orbitals and give rise to the peaked DOS around the Fermi energy. To describe the low-energy subspace around the Fermi energy we construct an effective three-orbital Wannier model, see also Sec.~\ref{sec:Method}. The left panels in Fig.~\ref{fig:dft_bands_dos} show that the resulting Wannier construction (without SOC) resembles the DFT bands and the total DOS to a very high accuracy. In order to include the effect of SOC, we add a local t$_{2g}$-only SOC term, see Sec.~\ref{sec:Method}, to the Wannier Hamiltonian. We find that a SOC strength of $\lambda=\SI{0.3}{eV}$ describes the DFT band structure and DOS calculated in the presence of SOC reasonably well, see right panels in Fig.~\ref{fig:dft_bands_dos}.
	Slight discrepancies between the Wannier model and DFT in the case including SOC are the result of neglecting possible $k$-dependent terms of the SOC.
	
	Without SOC, an important feature of the electronic structure of BaOsO$_3$ is a vHs exactly at the Fermi energy; notice the flat band close to the $X$ point in the top left panel of Fig.~\ref{fig:dft_bands_dos}. SOC (right panels) leads to a splitting of the vHs into two singularities, one below the Fermi level with dominant $j=3/2$ character and one above the Fermi energy with $j=1/2$ character. Although SOC does not impact the overall band width of the t$_{2g}$ states, we will see in the following that the splitting of the vHs has a profound impact on electronic correlations. Further, the SOC strength of $\lambda=\SI{0.3}{eV}$ is comparable to the Hund's coupling ($J_H\sim\SI{0.3}{eV}$), and thus Hund's-metal physics stands in competition with a SOC-driven polarization of the system.
	
	\subsection{Spectral function and correlated band structure}
	\label{sec:spec}
	
	\begin{figure}[t]
		\centering
		\includegraphics[width=\linewidth]{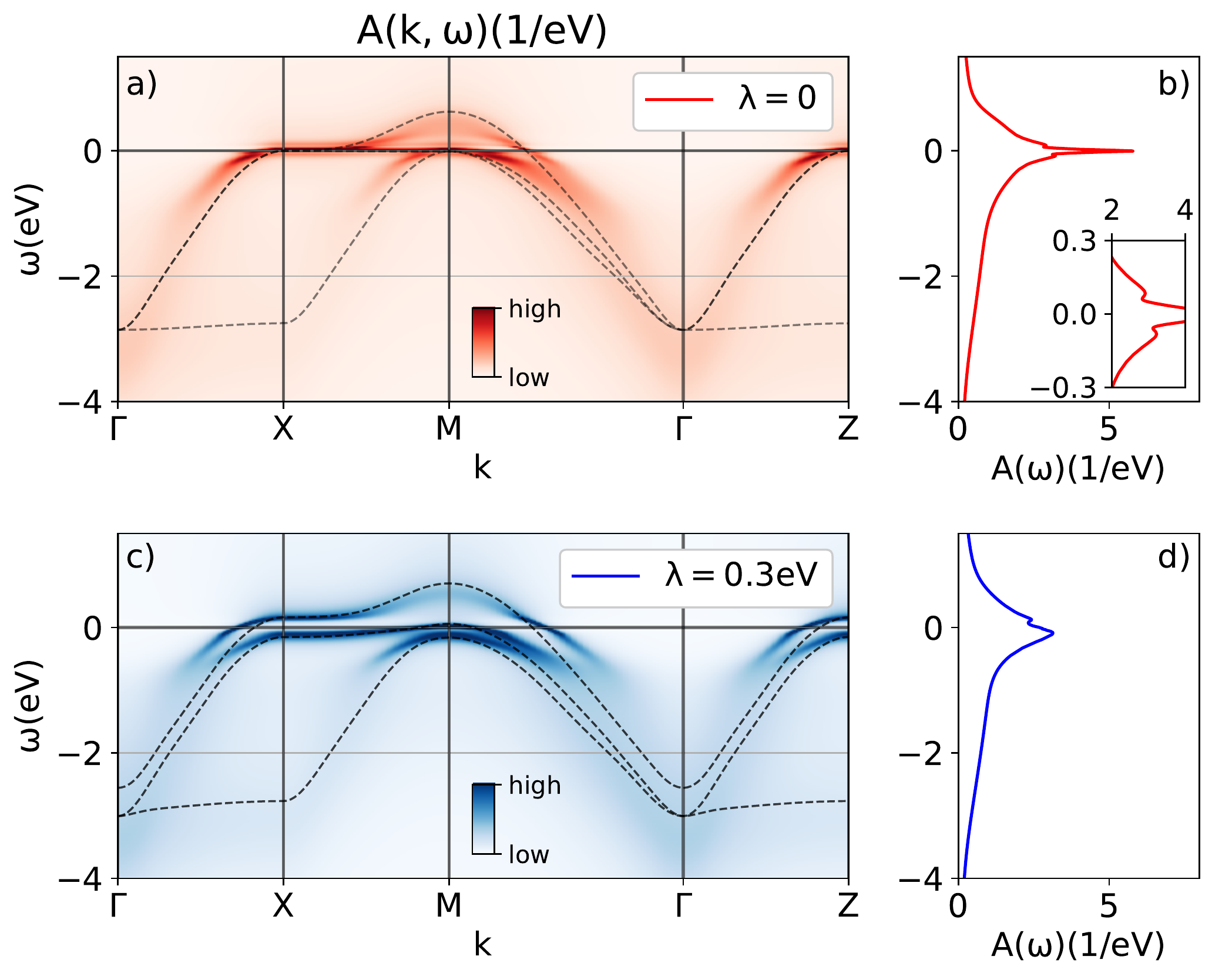}
		\caption{Momentum resolved spectral function for $U=\SI{2.55}{eV}$ and $J_H=\SI{0.27}{eV}$ along a high-symmetry path through the Brillouin zone without (a) and with (c) SOC together with the respective momentum-integrated spectral functions (b, d). The SOC strength is denoted as $\lambda$. The dashed lines in panel (a) and (c) correspond to the bare bands of the Wannier model. The inset in panel (b) shows a low-energy zoom of the spectral function.}
		\label{fig:mom_and_loc_specs}
	\end{figure}

	We now turn to the discussion of the correlated spectral function obtained with DMFT, shown in Fig.~\ref{fig:mom_and_loc_specs}. At physically realistic parameters ($U=\SI{2.55}{eV}$, $J_H=\SI{0.27}{eV}$, $\lambda=\SI{0.3}{eV}$) \BOO{} is a moderately correlated metal with strongly renormalized quasiparticle bands at low energies, and substantially incoherent states at higher energies. Without SOC the vHs is still present at the Fermi level and its splitting with SOC is relatively unaffected by electronic correlations. The calculated mass enhancement of 2.3 (with SOC) is in excellent agreement with the experimentally determined specific heat enhancement of 2.2~\cite{Shi2013}. We mention in passing that this good agreement with experiments gives us confidence in our choice of interaction parameters for \BOO{}. The overall band width of the correlated spectral function is roughly comparable to the one of the non-interacting model; this is commonly found in Hund's metals.
	
	Intriguingly, in the case without SOC we observe two side peaks, one on the unoccupied side of the spectrum at about \SI{0.1}{eV} and one at the same energy (\SI{-0.1}{eV}) on the occupied side, see panel (b) and inset of Fig.~\ref{fig:mom_and_loc_specs}. These peaks cannot be traced back to features already present in the non-interacting DOS. This is quite interesting, as recently the emergence of a side peak was identified as a characteristic feature of Hund's metals both in model studies~\cite{Kugler2019,Stadler2019,Horvat2019,Horvat2017} and in materials calculations~\cite{Stricker2014,Wadati2014,Kugler2020,Karp2020}. However, in these works a Hund's-metal side peak is either present on the occupied side of the spectrum for systems with less then half-filled shells, or on the unoccupied side for more than half-filled shells, but never simultaneously on both sides. In calculations for \BOO{} without Hund's coupling ($J_H=0$) we do not observe side peaks no matter if SOC is considered or not (spectral function not shown). Thus, the nature of the observed peaks is likely related to Hund's physics.

	On the other hand, the side peaks also vanish for $\lambda=\SI{0.3}{eV}$ and $J_H=\SI{0.27}{eV}$; note that the peaks visible in the corresponding spectral function (panel (d) of Fig.~\ref{fig:mom_and_loc_specs}) are already present in the non-interacting DOS (panel (d) of Fig.~\ref{fig:dft_bands_dos}), as they are a direct result of the splitting of the vHs due to SOC. To understand if the presence of the vHs at the Fermi energy is necessary for the emergence of the side peaks at finite $J_H$, we carry out additional calculations using a featureless semi-circular DOS and a band width tuned to yield the same mass enhancement as in \BOO{}, see Appendix~\ref{app:bethe}. We do not observe the side peaks in these calculations, and thus conclude that the presence of the vHs at the Fermi energy is indeed of key relevance. In contrast to other Hund's metals, the DOS of \BOO{} without SOC (see Fig.~\ref{fig:dft_bands_dos} panel (c)) is very symmetric between -0.5 and \SI{0.5}{eV} due to the vHs located right at the Fermi level. With the inclusion of SOC this symmetry in the spectral function is lost. Here, we can only speculate that the symmetry of the DOS supports both a peak on the occupied as well as the unoccupied side of the spectrum, but we suggest a more detailed study of the side peaks - probably on the level of simpler models - for future work.
	
	Additionally, we find that the material is much stronger correlated when SOC is switched off, i.e. the mass enhancement is about 50\% larger without SOC. This suppression of Hund's physics in \BOO{} with the inclusion of SOC might be an additional reason why the side peaks are only present in the case without SOC. To this end, it is necessary to disentangle the effects of the vHs in \BOO{} from the impact of Hund's physics. We will take a separate look at those two factors in the following sections, which will also help us to gain a deeper understanding of their interplay.
	
		\subsection{SOC and van-Hove singularity}
		\label{sec:vHs}

	\begin{figure}[t]
		\centering
		\includegraphics[width=0.9\columnwidth]{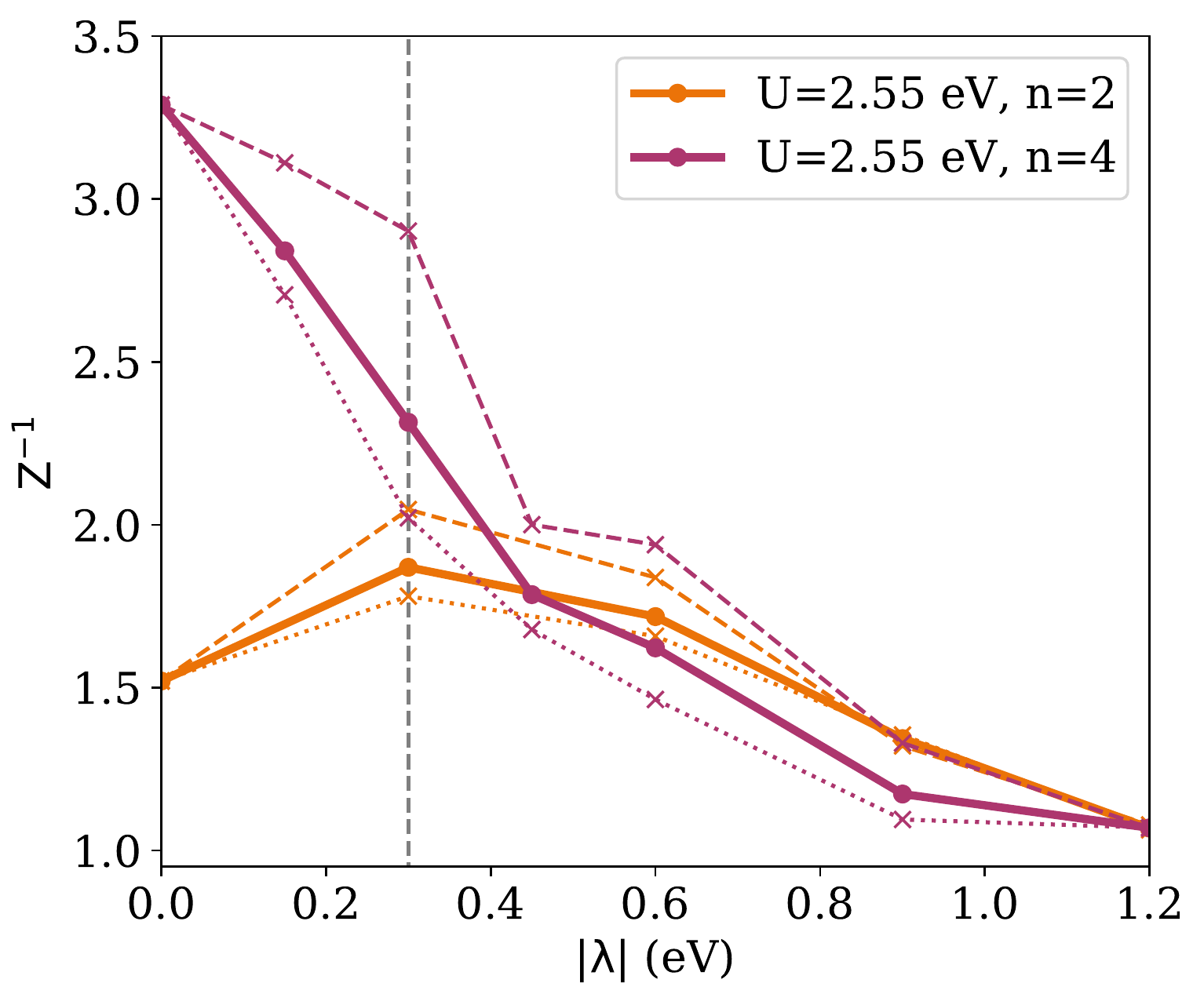}
		\caption{Mass enhancement $Z^{-1} = m^*/m$ as a function of
			the absolute value of  $\lambda$ for
			t$_{2g}$ fillings of $n=2$ (orange lines) and $n=4$ (purple lines) electrons with interaction
			parameters $U=\SI{2.55}{eV}$ and $J_H=\SI{0.27}{eV}$. In
			order to obtain a band insulating state also in the case $n=2$ (at large $\lambda$) we set $\lambda_{n=2}=-\lambda_{n=4}$. The dotted and dashed lines
			correspond to mass enhancement of the $j=3/2$ and $j=1/2$ bands
			respectively, while the thick solid line show the mass enhancement in the cubic basis. The dashed vertical line indicates the physical value of $\lambda$. }
		\label{fig:influence_vanHove}
	\end{figure}
	
	We have already discussed that the inclusion of SOC reduces the mass enhancement of BaOsO$_3$. In Fig.~\ref{fig:influence_vanHove} we show this effect in more detail as a function of the SOC strength $\lambda$  (purple lines).  Starting from vanishing SOC ($\lambda = 0$) with a high mass enhancement of 3.3, electronic correlations decrease with increasing SOC.
	In general the $j=1/2$ bands are more correlated than the $j=3/2$ bands. We attribute this to the higher degeneracy of the $j=3/2$ bands, resulting  in those states being farther away from half-filling (smaller active space). The reason behind this is that with increasing SOC the $j=1/2$ states empty out and pass through half-filling, while the $j=3/2$ states get filled and consequently move further away from the point of half-filling.
	
	Ultimately, in the limit of large $\lambda$ the mass enhancement is expected to vanish ($m^*/m$=1), because the system is then fully polarized and has already a gapped non-interacting DOS. Independent of the interaction strength or specific band structure details, like the vHs, the states of this band-insulating system are either occupied or empty, and thus dynamic correlations cease to play a role. The observation that $m^*/m$ approaches 1 in the limit of large SOC was also made by Triebl et al.~\cite{Triebl2018}, who investigated SOC in a three
	orbital model with a featureless semi-circular DOS. Unlike in the case of \BOO{}, their results show that $m^*/m$ stays essentially constant with increased SOC in the parameter range for which the system is still a metal. We infer that the decline in correlation before the band-insulating limit is reached in our calculations is due to changes in the non-interacting DOS with increasing SOC strength. We present the non-interacting DOS as a function of $\lambda$ in Appendix~\ref{app:DOS}.
	
	The impact of the SOC on band-structure details has also immediate implications for electronic correlations in the physical parameter regime. In the case of \BOO{}, the splitting of the vHs due to SOC leads to a decrease of the non-interacting DOS at the Fermi level and in turn to a decrease in electronic correlations, as e.g. argued in Refs.~\cite{Karp2020, Kugler2020}. 
	
	To obtain a more definite proof of the importance of the vHs at small SOC values, we carry out a numerical experiment: We consider \BOO{} with 4 holes instead of 4 electrons in the t$_{2g}$ orbitals and an opposite sign of the SOC strength $\lambda$, such that the SOC still favours a van-Vleck insulating state. With this artificial reduction to an occupation of only 2 electrons, we create a system that is effectively still at the same Hund's-metal filling, but it does not have a vHs in the vicinity of the Fermi level. In order to obtain an occupation of 2 electrons we adjust the chemical potential, but otherwise keep the band structure fixed.
	
	Unlike at the original filling, the problem with 2 electrons has a more than 2 times smaller mass enhancement of about 1.5 at $\lambda=0$, see Fig.~\ref{fig:influence_vanHove} (orange lines). This is very much in line with our previous findings on \SMO{} and \SRO{}~\cite{Karp2020}.
	Electronic correlations in the problem with 2 electrons also show less dependence on $\lambda$ and the results are overall more similar to the ones obtained for a semi-circular DOS~\cite{Triebl2018}. The weak dependence on $\lambda$ can be explained by the fact that the non-interacting DOS at the Fermi energy does barely change for $\lambda \leq \SI{0.9}{eV}$, as shown in Appendix~\ref{app:DOS}. Surprisingly, we see an initial increase in the mass enhancement when increasing the SOC
	to \SI{0.3}{eV}. We relate this observation to the fact that in the 2-electron system the integer occupancy of both the $j=1/2$ and $j=3/2$ subspace is crossed around $\lambda=\SI{0.3}{eV}$.
	As SOC increases further the differences between the calculations with 2 and 4 electrons become less pronounced and at both fillings the mass enhancement vanishes in the limit of large SOC. 
	The small influence of SOC on the mass enhancement due to a weak dependence of the DOS around the Fermi level in the 2-electron system is a strong evidence that the vHs is responsible for the enhanced correlations in \BOO{} for $\lambda \leq \SI{0.3}{eV}$. Our results clearly demonstrate the dramatic impact band-structure aspects can have on electronic correlations.
	
	\subsection{SOC and Hubbard/Hund's physics}
	\label{sec:SOC}
		\begin{figure*}[t]
		\centering
		\includegraphics[width=1.0\linewidth]{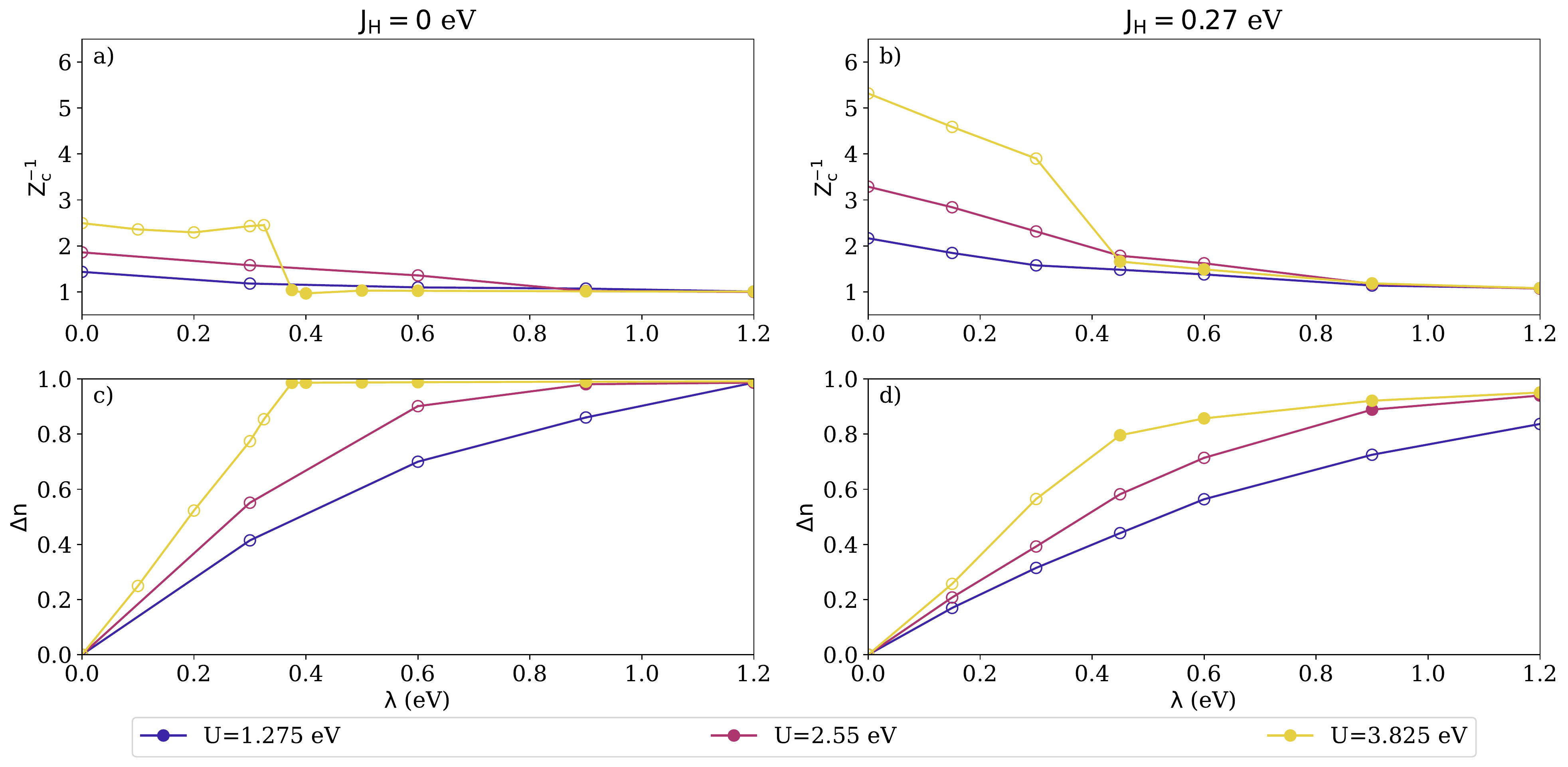}
		\caption{Diagonal elements of the mass enhancement $Z_c^{-1} = m^*/m$ in the cubic basis (a, b) and polarization
			$\Delta n := n_{j=3/2}-n_{j=1/2}$ (c, d) for different values of
			$U$. The occupation numbers used
			for the polarization were taken directly from the
			ground state. Open and full circles display metallic and insulating states respectively.}
		\label{fig:mass_enhancement}
	\end{figure*}
	
	In order to gain an understanding of the role of Hund's physics in \BOO{} we computed the mass enhancement, Fig.~\ref{fig:mass_enhancement} (top panels), and the polarization $\Delta n = n_{j=3/2}-n_{j=1/2}$ of the states in the $j$ basis, Fig.~\ref{fig:mass_enhancement} (bottom panels), for a broad range of $\lambda$ and interaction parameters. Starting with $J_H=0$, Fig.~\ref{fig:mass_enhancement} (left panels), the mass enhancement is only weakly influenced by SOC and stays approximately constant up to a critical SOC value at which it abruptly drops to 1. This is best seen in the calculations for $U=\SI{3.825}{eV}$ (yellow lines), with a critical $\lambda$ of about $\SI{0.4}{eV}$, which is quite close to the physical SOC value. However, for a more realistic interaction of $U=\SI{2.55}{eV}$ already $\lambda>\SI{0.8}{eV}$ would be necessary (at $J_H=0$) to reach the insulating regime.
	
	At the same time, the system gets perfectly polarized at the critical $\lambda$ value, corresponding to fully occupied $j=3/2$ states and empty $j=1/2$ states. At a finite SOC strength and $J_H=0$, the material always transitions to a band-insulating van-Vleck regime, but the critical SOC strength depends on $U$. The lower critical $\lambda$ observed with increasing $U$ points to a polarizing effect of the Hubbard interaction, that is, the susceptibility to a $j$-splitting is enhanced. In practice, this is directly visible in the real part of the self-energy, and has been interpreted as a correlation-driven enhancement of the effective SOC~\cite{Liu2008,Kim2018,Tamai2018,Linden2020}. For a more detailed discussion of this matter, we refer the reader to the next section.
	
	Turning now to $J_H=\SI{0.27}{eV}$ (Fig.~\ref{fig:mass_enhancement} right panels), our overall observation is a substantially larger mass enhancement for $\lambda\leq \SI{0.3}{eV}$, which illustrates the Hund's-ness of \BOO{}. Without SOC, the Hund's coupling leads to a doubling of the mass enhancement at $U=\SI{3.825}{eV}$, and still yields an increase by 2/3 for the physical $U$ value of $\SI{2.55}{eV}$ in comparison to the calculations for $J_H=0$. The magnitude of the enhancement is comparable to the effect of the vHs, see Sec.~\ref{sec:vHs}. In summary, the unusually large mass enhancement of 3.3 at $\lambda=0$ for a 5d material, can be traced back - in equal parts - to Hund's physics and the presence of the vHs at the Fermi energy.
	
	In comparison to 3d/4d Hund's metals~\cite{Mravlje2011,Karp2020,Wadati2014,Yin2011, Deng2019}, the increase in correlations due to $J_H$ is less pronounced in \BOO{}. Nevertheless, our results show that Hund's physics persists to affect correlations even at the physical SOC value ($\lambda=\SI{0.3}{eV}$), although the enhancement here is only about 50\%. This is in part due to $J_H$ acting against the orbital polarization and resisting the transition to a band insulator, which is well apparent in the polarization (Fig.~\ref{fig:mass_enhancement} bottom panels): In the case of $J_H=0$ we find that a full polarization ($\Delta n = 1$) is approached rapidly and even attained, while in the case of finite $J_H$ the overall behavior is smoother and a full polarization is not attained in the explored parameter regime. Instead, with increasing SOC the system gets continuously more and more polarized without showing a clear transition to a van-Vleck band insulator. The smoothness of this transition at finite $J_H$ seems to be a generic feature that is independent of the band structure~\cite{Triebl2018}.
	
	In essence, the fact that there is a competition between Hund’s coupling and SOC can be understood by considering the atomic limit. Namely, the terms of the Hamiltonian proportional to $J_H$ favor a state with maximum spin $S$ and angular momentum $L$, while the polarized system with fully occupied $j=3/2$ states corresponds to a total angular momentum $J_\mathrm{tot}=0$
	state. This is not a contradiction, since there exists a total angular momentum $J_\mathrm{tot}=0$ state for the maximal values $S=1,L=1$ in the case at hand. However, computing the overlap of the
	$L=1,S=1, J_\mathrm{tot}=0$ state with the one having 4 electrons in the $j=3/2$ and 2 holes in the $j=1/2$ states, yields only $\sqrt{2/3}<1$. Therefore the fully polarized state also has to contain contributions from the $L=0, S=0, J_\mathrm{tot}=0$ state, which is the only other atomic state yielding $J_\mathrm{tot}=0$. This means that the polarized state favored by SOC contains contributions that are penalized by Hund's coupling, leading to a competition between the two as long as $J_H$ and $\lambda$ are finite.
	
	Indeed, our results show that Hund's physics is dominant for $\lambda \ll J_H$, apparent by the difference in the mass enhancement found between $J_H=0$ and $J_H=\SI{0.27}{eV}$ for $\lambda\leq\SI{0.3}{eV}$. The effect of the SOC dominates, on the other hand, for $J_H \ll \lambda$, visible e.g. in the mass enhancement approaching 1 in the limit of large $\lambda$. With $J_H\sim\lambda$, SOC and Hund's coupling are equally important in \BOO{} and neither can be neglected to correctly describe the physics of this material. The competition between SOC and Hund's physics also allows for a smooth crossover between the Mott and band-insulating state, as we will show in the next section.

	\subsection{Phase diagram}
	\label{sec:Phases}
	\begin{figure*}[t]
		\centering
		\includegraphics[width=\linewidth]{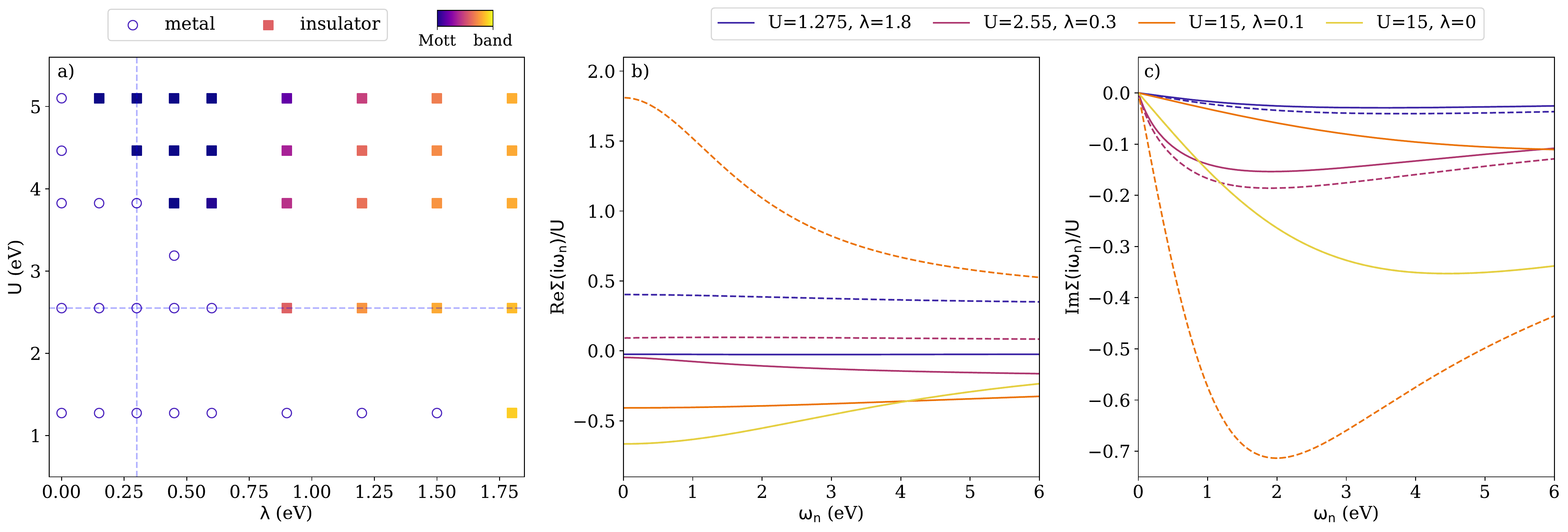}
		\caption{Phase diagram in the $U - \lambda$ plane at constant $J_H=\SI{0.27}{eV}$ (a).
			We find two different phases in the parameter regime considered, namely a metallic phase (open circles), and an insulting phase (squares). The character of the insulating states (Mott-like or band-like) is encoded in the symbol colors (see main text). The colors are scaled linearly between $\chi=0.5$ and \SI{0}{eV}, while for $\chi\geq\SI{0.5}{eV}$ the color is set to the one for $\chi=\SI{0.5}{eV}$. The dashed lines indicate the physical values of $\lambda$ (vertical line) and $U$ (horizontal line). 
			Real (b) and imaginary (c)
			part of the self-energy for $J_H=\SI{0.27}{eV}$ and selected values of $U$ and $\lambda$ (in eV). The solid and dashed lines correspond to the $j=3/2$ and $j=1/2$ bands, respectively. In order to show all self-energies on a comparable scale we shifted the real parts by the chemical potential and rescaled both imaginary and real part by $1/U$.}\label{fig:SelfEs}
	\end{figure*}

	Having gained insights on the role of SOC, Hund's coupling and the vHs, we will now place \BOO{} within a paramagnetic phase diagram in the $U-\lambda$ plane.
	In the left panel of Fig.~\ref{fig:SelfEs} we present the paramagnetic phase diagram, highlighting the regions where we identify
	metallic (circles) or insulating (squares) behaviors. The insulating phases are distinguished from the metallic ones using the value of the local Green's function at
	$\omega_n \rightarrow 0$, which corresponds to the momentum-integrated spectral function at the chemical potential.  We note that our phase diagram and the metal-insulator boundary is consistent with an unpublished calculation by Dai~\cite{dai_kitp_talk2011}. Both $U$ and $\lambda$ stabilize the insulating phase, most effectively when both act together, e.g. an insulator is already found for $U\sim\SI{3.5}{eV}$ and $\lambda=\SI{0.6}{eV}$. \BOO{} (indicated by the dashed blue lines) still falls in the metallic region, but is not far from the boundary to the insulating state.
  
  	We will now undertake a further characterization of the insulating region, as the Hubbard interaction $U$ and SOC can lead to distinct phases. At $\lambda=0$, a Mott insulating phase is expected for sufficiently large $U$. The Mott-insulating phase is fundamentally different to a band insulator, as it is a phenomenon driven by electronic correlations. On the contrary, the mechanism behind the band-insulating van-Vleck phase is the opening of a gap between the $j=3/2$ and $j=1/2$ states due to strong SOC, which is at its core a single particle effect.
    When $\lambda$, $U$ and $J_H$ are present, there is a smooth crossover between the two insulating phases, shown with the color coding of the squares in the left panel of Fig.~\ref{fig:SelfEs}.
  	To determine the color coding we use $\chi(U,J_H,\lambda)=\left\vert \text{min}_{j, \omega_n}\text{Im} \Sigma_j\left(i\omega_n\right) \right\vert$, where $\text{Im} \Sigma_j(i\omega_n)$ is the imaginary part of the local self-energy for orbital angular momentum $j$, which is a suitable measure for the size of dynamic correlation.

	To illustrate the distinction of Mott-like and band-insulating-like regimes, we discuss the Matsubara self-energies for several characteristic cases in panels (b, c) of Fig.~\ref{fig:SelfEs}. First, $U=\SI{1.275}{eV}$ and $\lambda=\SI{1.8}{eV}$ shows a typical
	band-insulator behavior. The imaginary parts are small and the real parts are approximately constant. The correlations do have an important effect nevertheless: they induce a splitting of the $j=1/2$ and $j=3/2$ components of the self-energy and thus enhance the orbital polarization. Second, we consider a typical Mott insulator without SOC, using $U=\SI{15}{eV}$ and $\lambda=0$. The imaginary part of the self-energy is much larger and strongly frequency dependent in comparison to the first case and demonstrates the importance of dynamic electronic correlations in a Mott insulator. Notice that the $j$ states are degenerate for $\lambda=0$.
	
	Our third case, $U=\SI{15}{eV}$ and $\lambda=\SI{0.1}{eV}$, corresponds to a Mott
	insulator at small but finite SOC. The SOC breaks the spin symmetry, allowing for a further polarization of the system by the overwhelmingly large Hubbard repulsion. This is apparent in the gap of the real parts, which is far larger than e.g. in the case of $U=\SI{1.275}{eV}$ and $\lambda=\SI{1.8}{eV}$. Inspecting the imaginary part we see that the degeneracy of the $j=3/2$ and the $j=1/2$ bands is now lifted. There are strong correlations mainly in the $j=1/2$ bands, while in the $j=3/2$ bands they are much less pronounced. This is due to the effect of a higher band degeneracy in the latter, as explained in Sec.~\ref{sec:vHs}. The strong frequency dependence and large absolute values in the $j=1/2$ imaginary part indicate that the system is actually far from a band insulator. This is not obvious, as one might have expected that the polarizing effect of the large $U$ would immediately drive the system into a band-insulating regime for $\lambda\neq0$.

	Finally, for the parameters we deem realistic, we find a substantial imaginary part of the self-energy, confirming the sizable dynamic correlations in \BOO{}. In the literature on \SRO{}, the splitting found in the real part of the self-energy has been interpreted as a correlation-driven enhancement of the effective SOC~\cite{Kim2018,Tamai2018,Karp2020,Linden2020,Triebl2018}. In contrast to \SRO{}, the real part is however strongly frequency dependent in \BOO{}, which does not allow for such an interpretation in an unambiguous way. Following the arguments of Triebl et al.~\cite{Triebl2018} we use the definition $\lambda_\mathrm{eff}:=\lambda + \frac{2}{3}\Delta(\text{Re}\Sigma_j(i\omega_n \rightarrow 0)$. 
	In this low frequency limit the splitting between the components of the real part of the self-energy is about \SI{0.34}{eV}, which yields an effective correlation-enhanced SOC of about \SI{0.53}{eV}. This means that the bare SOC is effectively enhanced by roughly a factor of 1.8 (in the low frequency limit), which is slightly less than the factor of 2 found for \SRO{} and \SMO{}~\cite{Kim2018,Tamai2018,Karp2020,Linden2020}.
	
	Interestingly, $\lambda_\mathrm{eff}$ is not directly visible in the correlated spectral function, see Fig.~\ref{fig:mom_and_loc_specs} (c), where the splitting of the vHs due to SOC is roughly the same as in DFT. It was elaborated in Ref.~\cite{Tamai2018} that the effective splitting of \emph{quasiparticle} bands is, however, given by $\sim Z\lambda_\mathrm{eff}$, which is about \SI{0.23}{eV} in \BOO{}. This value is in good agreement with the splitting of the bands at the $X$ point of roughly \SI{0.25}{eV} (see Fig.~\ref{fig:mom_and_loc_specs} (c)). 

\section{Conclusion}
In this paper we discussed the electronic structure and correlation
effects in \BOO{}. We mapped out a phase diagram in the plane of the
Hubbard interaction $U$ and SOC strength $\lambda$. For
realistic interaction and SOC parameters we find that \BOO{} is a moderately correlated metal;
consistent with experiments. In \BOO{} the physical (bare) SOC of about $\SI{0.3}{eV}$ is not strong enough to induce a van-Vleck insulating regime, in contrast to e.g. the 5d iridate NaIrO$_3$~\cite{du2013,Bremholm2011}. Due to a nominal filling of 4 electrons in the 3 Os-t$_{2g}$ orbitals, \BOO{}
falls into the Hund's-metal regime, but has a 3-4 times higher SOC than the well-studied 3d/4d
Hund's metals, like iron-pnictides and ruthenates. Additionally, the non-interacting DOS (without SOC) of \BOO{} has a van-Hove singularity at the Fermi energy. This is another similarity
to the ruthenates Sr$_2$RuO$_4$~\cite{Mackenzie1996b,Maeno1997,Bergemann2003,Stricker2014,
	Behrmann2012, Tamai2018, Deng2016, Veenstra2014,
	Zingl2019,Sarvestani2018,Lee2020,Zhang2016,Kim2018,Strand2019,Aiura2004,Mravlje2011,Deng2019,Lee2020,Kugler2020, Karp2020} and BaRuO$_3$~\cite{han2016,dasari16}, where the van-Hove singularity is known to strongly impact electronic correlations and their physical properties. 

To understand the influence of SOC on electronic correlations in \BOO{} we studied various SOC strengths. Our calculations at small SOC show that electronic correlations in \BOO{} are indeed driven by the Hund's coupling and the proximity of the vHs close to the Fermi energy.
Increasing SOC leads to a nontrivial decrease of electronic correlations, as (a) the importance of Hund's coupling diminishes due to the competition of the atomic states favored by SOC and Hund's coupling (cf. Sec.~\ref{sec:SOC}), and (b) the SOC splits the vHs into two parts reducing the DOS at the Fermi energy (cf. Sec.~\ref{sec:vHs}). Here the effect (a) is generic and will apply to other oxides with degenerate  t$_{2g}$ orbitals at $n=4$ filling irrespectively of the precise shape of the DOS, while consequence (b) of SOC is non-generic and should be considered as pertaining to materials that have a van-Hove singularity at the Fermi level in the case of vanishing SOC.
 
Our work demonstrates that electronic correlations in \BOO{} are governed by a complex interplay of Hund's physics, SOC and details of the band structure (vHs). By studying all these factors and their impact on electronic correlations, we believe that our results contribute towards a holistic understanding of quantum materials and their fascinating properties.\\

\section*{Acknowledgments}
We thank A.~Georges, A.~J.~Millis and O.~Parcollet for fruitful
discussions. We gratefully acknowledge feedback from
J.~Karp and M.~Aichhorn. J. M. is supported by the Slovenian Research Agency (ARRS) under Program No. P1-0044, J1-1696, and J1-2458. M.B., M.G. and U.S. acknowledge support by the Deutsche Forschungsgemeinschaft (DFG, German Research Foundation) under
Germany's Excellence Strategy-426 EXC-2111-390814868 and by Research
Unit FOR 1807 under Project No. 207383564. We acknowledge funding through
the ExQM graduate school. M.B., M.G. and U.S. thank
the Flatiron Institute for its hospitality. The Flatiron Institute is
a division of the Simons Foundation. 


\appendix

\section{Interaction Hamiltonian} \label{app:method}

We use Hubbard-Kanamori on-site interactions~\cite{Kanamori1963}
\begin{multline*}
H = U \sum_l n_{l \uparrow}n_{l \downarrow} 
+ \sum_{l<l', \sigma}[U'n_{l\sigma}n_{l'\Bar \sigma}
+ (U' - J_H)n_{l\sigma}n_{l'\sigma} \\
- J_Hc^\dagger_{l\sigma} c_{l\Bar \sigma} c^\dagger_{l' \Bar \sigma} c_{l' \sigma}]
- J_H \sum_{l<l'} [c^\dagger_{l\uparrow}c^\dagger_{l\downarrow}c_{l'\uparrow}c_{l'\downarrow} + h.c.]
\end{multline*}
with $l\in\{xy, xz,yz\}$ and $U' = U - 2J_H$. The values used for $U$ and $J_H$ are provided in the main text.

The local SOC term we consider is given by
\begin{equation*}
	H_{SOC}=\frac{\lambda}{2} \sum_{mm^\prime} \sum_{\sigma \sigma^\prime} ({\bf l}_{mm^\prime} \cdot \bm{ \sigma}_{\sigma \sigma^\prime}) c^\dagger_{m\sigma} c_{m^\prime \sigma^\prime}
\end{equation*}
where $\lambda$ is the SOC strength, ${\bf l}$ are the t$_{2g}$-projected angular momentum matrices and $\bm{\sigma}$ is a vector of Pauli matrices (cf. Refs.~\cite{Linden2020,Karp2020}). Note that the orbitals indicated by $m$ and $m^\prime$ are the t$_{2g}$ orbitals in the cubic basis, i.e. $m,m^\prime \in \{xy,xz,yz\}$.

\section{MPS-based solver}
\label{app:mps}
For DMFT we use the MPS-based impurity solver in imaginary time, introduced in Ref.~\cite{wolf15iii} and successfully applied in the context of DFT+DMFT in Refs.~\cite{Linden2020,Karp2020}. Details concerning the method are given in Refs.~\cite{wolf15iii,Linden2020}. All calculations are performed using the \textsc{SyTen} toolkit~\cite{hubig17, systen}.

As system size we use $L_b=6$ bath sites per spin and orbital yielding a total of $L_{tot}=21$ sites for calculations with SU(2) symmetry and a total of $L_{tot}=42$ sites without. We use a frequency grid corresponding to Matsubara frequencies at a (fictitious) inverse temperature $\beta_{\text{fict}}=\SI{200}{ eV^{-1}}$.

In general it is important to implement as many symmetries as possible, in order to guarantee convergence to the correct ground state. Therefore, without the inclusion of SOC and at $J_H\neq 0$, we use five quantum numbers: the occupation parity of each orbital, the particle number, and the spin. In the case of $J_H=0$, the occupation number of every band is conserved, thus we implemented those instead of the parity. In ground state searches we allow for bond dimensions up to 4096, however in most parameter regimes this limit is not reached. For the time evolution we use the time-dependent-variational-principle (TDVP)~\cite{haegeman11,haegeman16,paeckel2019} up to $\tau=\SI{100}{eV^{-1}}$ in steps of $\Delta \tau=\SI{0.05}{eV^{-1}}$, supplemented by linear prediction~\cite{barthel09} to extrapolate to larger times. We consider DMFT loops to be converged when the largest change in the hybridization function is below $10^{-3}$. Due to the threefold degeneracy of the bands in \BOO{} we only compute the time-evolution for one of the bands and determine all others by symmetry.\\

\begin{figure}[t]
	\centering
	\includegraphics[width=\columnwidth]{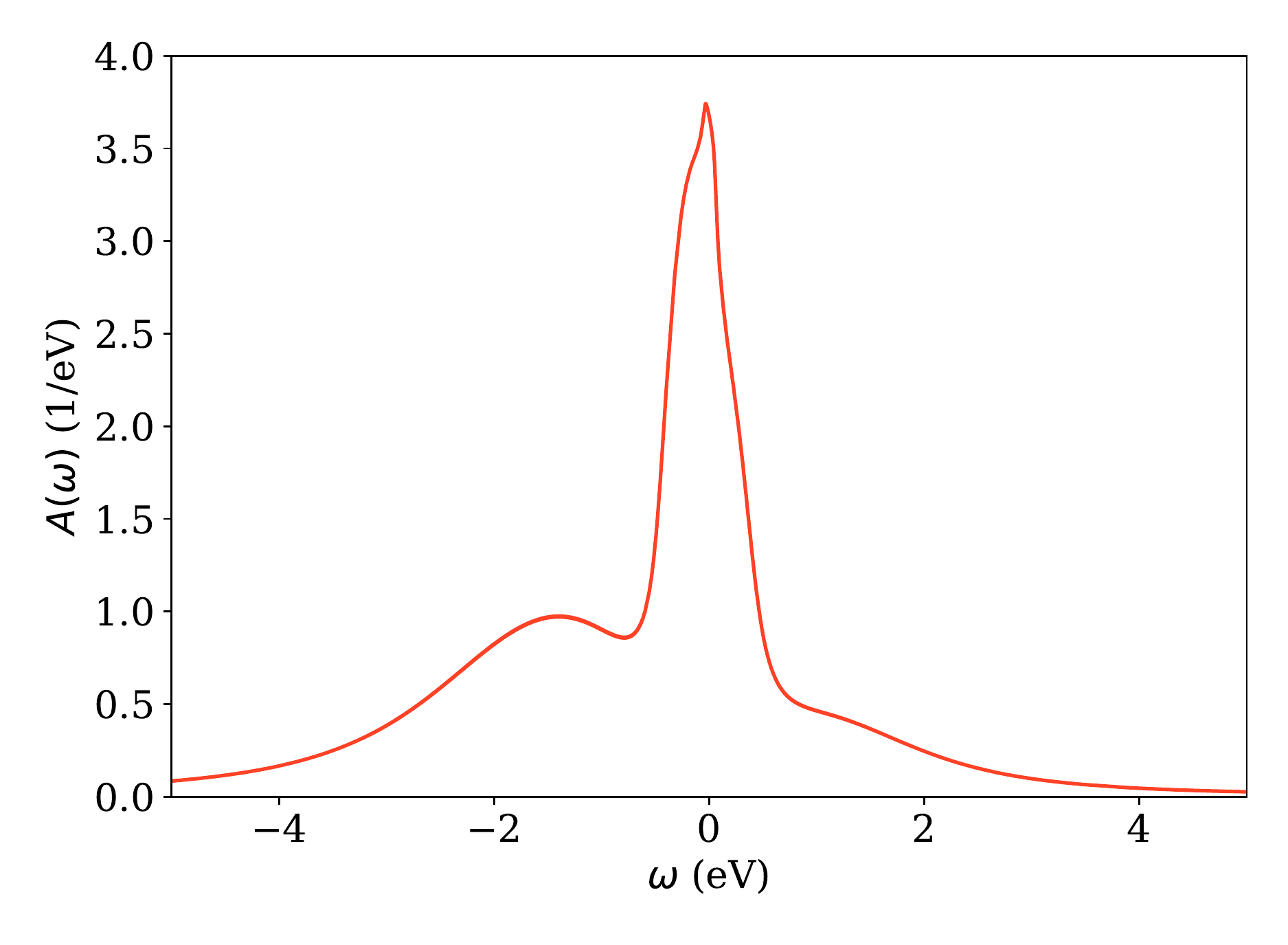}
	\caption{Local spectral function of a model calculation on a semi-circular DOS with $U=\SI{2.55}{eV}$, $J_H=\SI{0.27}{eV}$, $\lambda=0$ and half bandwidth $D=\SI{1}{eV}$. }
	\label{fig:bethe_spec}
\end{figure}

\begin{figure}[t]
	\centering
	\includegraphics[width=\columnwidth]{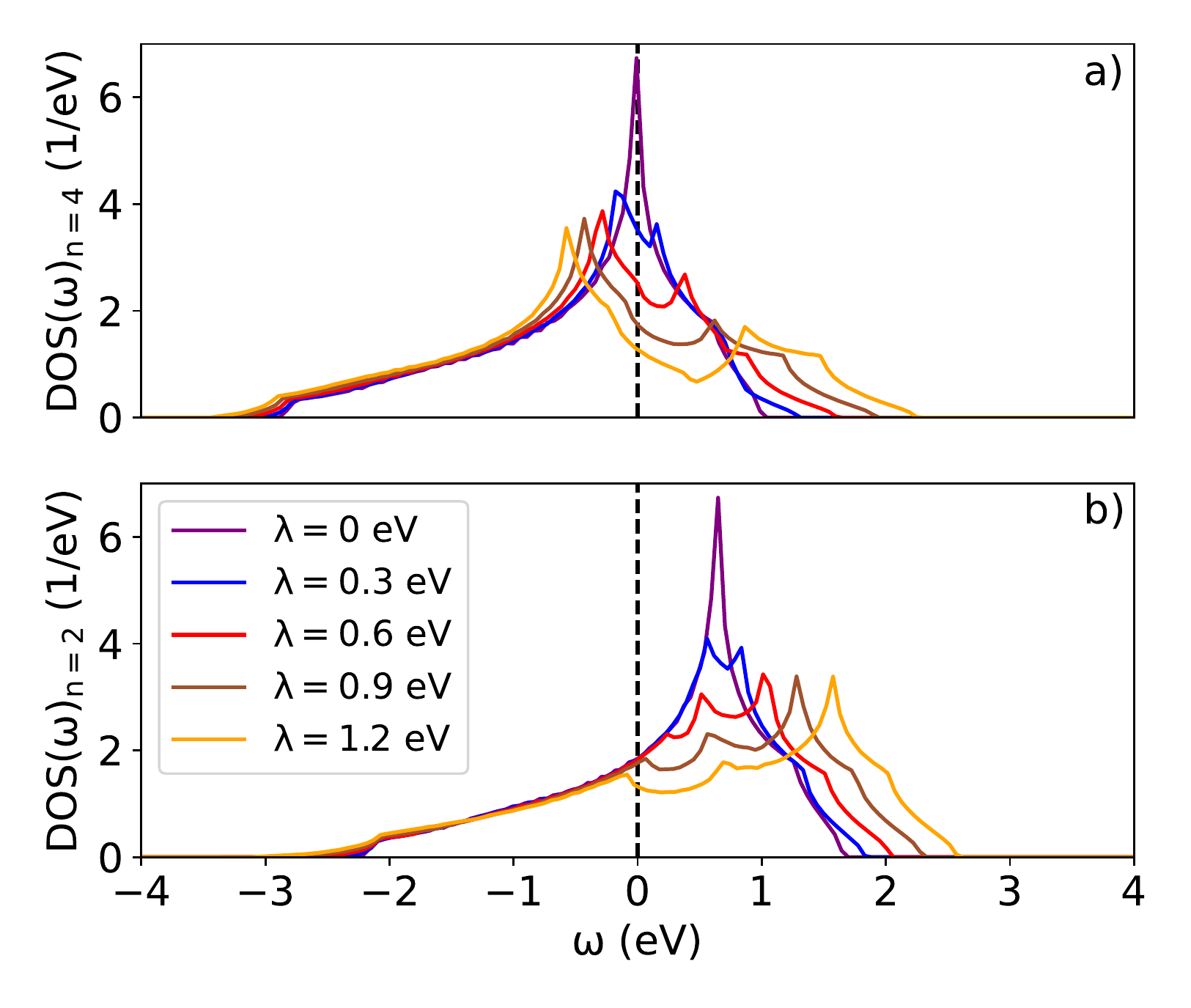}
	\caption{Non-interacting density of states (DOS) at a filling of 4 (a) and 2 electrons (b).
		For the system with 2 electrons we also flipped the sign of $\lambda$ (see text).}
	\label{fig:DOS}
\end{figure}

With SOC included, we use different quantum numbers since the spin SU(2) symmetry is broken. In the case of finite Hund's coupling our quantum numbers are the particle number and the $z$-component of the total angular momentum in the $j$ basis. When the Hund's coupling is switched off ($J_H=0$), the occupation number in every band is conserved, which yields six different quantum numbers corresponding to the conservation of particle numbers in every band.

Even with the inclusion of SOC the Green's function is still diagonal in the $j$ basis due to the threefold degeneracy of the bands in \BOO{} and thus every band can be fitted separately. Furthermore a significant part of the degeneracy stays present in the system. Namely, the bands with a given $j$ are degenerate, which we use in order to compute the Green's functions only for one band per $j$ and determine the rest by symmetry.

\section{Model calculation}
\label{app:bethe}

In order to check whether the side peaks present in the local spectral function for $U=\SI{2.55}{eV}$, $J_H=\SI{0.27}{eV}$ and $\lambda=0$ (see Fig.~\ref{fig:mom_and_loc_specs} (b)) are due to the vHs located exactly at the Fermi level, we perform calculations on a semi-circular DOS with a half bandwidth of $D=\SI{1}{eV}$. We choose $D$ such that the mass enhancement of the model calculation ($Z^{-1}=3.30$) fits that of \BOO{} ($Z^{-1}=3.29$) for the given parameter set.
The corresponding local spectral function is shown in Fig.~\ref{fig:bethe_spec}. Apart from the quasi-particle peak that features some internal structure there are no clearly pronounced side peaks as observed for \BOO{}. This supports our interpretation that the side peaks found in \BOO{}  are a result of the interplay between $J_H$ and the vHs at the Fermi level.

\section{Density of states}
\label{app:DOS}
To support the discussion in Sec.~\ref{sec:vHs} we present in Fig.~\ref{fig:DOS} (panel a) the non-interacting DOS of \BOO{} for $\lambda$ values between 0 and \SI{1.2}{eV}. SOC leads to a splitting of the vHs and a decrease of the DOS around the Fermi energy with increasing SOC, which in turn results in a decrease of the mass enhancement, see Sec.~\ref{sec:vHs}. 

In panel (b) we show the non-interacting DOS at a filling of 2 electrons and an opposite sign of $\lambda$. As mentioned in the main text, using an opposite sign of $\lambda$ is necessary to obtain the same physics as in the case of 4 electrons. It is simple to illustrate this in the limit of large $\lambda$: In \BOO{} the SOC lowers the 4 j=3/2 states in comparison to the 2 j=1/2 states. In the limit of large $\lambda$ the system is a band insulator with full j=3/2 states and empty j=1/2 states. This would not be the case when using a positive sign for $\lambda$ at a filling of 2 electrons.
In order for the SOC to play a comparable role in the 2-electron system it is necessary to flip its sign. Then, the SOC lowers the j=1/2 states with respect to the j=3/2 ones and 
again a band insulating ground state is found in the limit of large $\lambda$ (full j=1/2 and empty j=3/2 states).

In the 2-electron system the vHs is positioned well above the Fermi energy in the unoccupied part of the spectrum. With increasing $\lambda$ we also observe a splitting of the vHs, but the direction of it is inverted compared to the 4 electron case, since we
flipped the sign of the SOC. In contrast, the opposite side of the spectrum is almost unchanged with increasing $\lambda$, at least for the parameter range we explored. At the Fermi energy we observe only a very weak $\lambda$ dependence of the DOS. This is the reason for the weak dependence of the mass enhancement on SOC in the 2-electron system, as argued in Sec.~\ref{sec:vHs}.

\bibliographystyle{apsrev4-1}
\bibliography{references.bib}

\end{document}